\documentclass[11pt,a4paper]{article}
\pdfoutput=1

\usepackage{jcappub}
\usepackage{multirow}
\usepackage{amssymb} 
\usepackage[normalem]{ulem}

\newcommand{\dd}{{\rm{d}}}

\newcommand{\be}{\begin{equation}}
\newcommand{\ee}{\end{equation}}

\newcommand{\kop}
{\mathfrak{K}}

\definecolor{darkred}{rgb}{0.5,0,0}
\definecolor{darkgreen}{rgb}{0,0.6,0}
\definecolor{darkblue}{rgb}{0,0,0.5}
\definecolor{mypurple}{RGB}{180,20,180}

\newcommand{\Hc}{\mathcal{H}}

\newcommand{\CLASS}{{\sc class}}
\newcommand{\CAMB}{{\sc camb}}

\newcommand{\inspire}[1]{\href{http://inspirehep.net/search?p=find+J+#1}
 {{\color{black}[{\color{blue} {\small in}SPIRE}]}}}
\newcommand{\book}[1]{\href{http://inspirehep.net/search?p=#1}
 {{\color{black}[{\color{blue} {\small in}SPIRE}]}}}

\newcommand{\inspired}[1]{\href{http://inspirehep.net/search?p=#1}
 {{\color{black}[{\color{blue} {\small in}SPIRE}]}}}

\newcommand{\HL}{H_{\rm L}}
\newcommand{\HT}{H_{\rm T}}

\newcommand{\nab}{\nabla}

\makeatletter
\newsavebox\myboxA
\newsavebox\myboxB
\newlength\mylenA

\newcommand*\mybar[2][0.75]{%
    \sbox{\myboxA}{$\m@th#2$}%
    \setbox\myboxB\null
    \ht\myboxB=\ht\myboxA%
    \dp\myboxB=\dp\myboxA%
    \wd\myboxB=#1\wd\myboxA
    \sbox\myboxB{$\m@th\overline{\copy\myboxB}$}
    \setlength\mylenA{\the\wd\myboxA}
    \addtolength\mylenA{-\the\wd\myboxB}%
    \ifdim\wd\myboxB<\wd\myboxA%
       \rlap{\hskip 0.5\mylenA\usebox\myboxB}{\usebox\myboxA}%
    \else
        \hskip -0.5\mylenA\rlap{\usebox\myboxA}{\hskip 0.5\mylenA\usebox\myboxB}%
    \fi}
\makeatother

\begin{document}

\title{Suitable Initial Conditions for Newtonian Simulations with Massive Neutrinos}

\date{\today}

\author[a]{Christian Fidler,}
\emailAdd{fidler@physik.rwth-aachen.de}

\author[a]{Alexander Kleinjohann}
\emailAdd{kleinjohann@physik.rwth-aachen.de}

\affiliation[a]{Institute for Theoretical Particle Physics and Cosmology (TTK), RWTH Aachen University, Otto-Blumenthal-Strasse, D--52057 Aachen, Germany.}

\abstract{
Initial conditions for cosmological N-body simulations are usually calculated by rescaling the present day linear power spectrum obtained from an Einstein-Boltzmann solver to the initial time employing the scale-independent matter growth function. For the baseline $\Lambda$CDM model, this has been shown to be consistent with General Relativity (GR) even in the presence of relativistic species such as photons. We show that this approach is not feasible in cosmologies with massive neutrinos and present an alternative method employing the Newtonian motion gauge framework.
}

\maketitle   

\flushbottom


\section{Introduction}
\label{Introduction}

According to the $\Lambda$CDM paradigm, the Universe consists of a cosmological constant $\Lambda$, cold dark matter (CDM), baryons, photons, and neutrinos with their dynamics governed by the coupled Einstein-Boltzmann equations. Solving these in full non-linearity is currently not feasible, but at sufficiently early times, when the fluctuations are still small, we may use linear cosmological perturbation theory (CPT; \cite{Kodama:1985bj,Malik:2008im,Villa:2015ppa}). The corresponding system of equations can be solved efficiently by Boltzmann codes such as \CLASS~\cite{class2} and \CAMB~\cite{Lewis:1999bs}. 

At later times, the density perturbations become significantly larger than their initial values and eventually enter the non-linear regime. Today's observed large-scale structure (LSS) stems from the full non-linear gravitational evolution of these small initial perturbations. In order to accurately capture the dynamics we need methods that go beyond CPT. A common solution is the usage of cosmological N-body simulations \cite{Teyssier:2001cp,Springel:2005mi,Hahn:2015sia}, which approximate the phase-space dynamics of matter to high accuracy using a large number of tracer particles. 

The initial conditions for such simulations could be set at a time when the linear approximation is still accurate, and calculated using an Einstein-Boltzmann solver. However, to simplify the problem of solving the Einstein-Boltzmann system in its full non-linearity, N-body simulations usually employ a Newtonian gravity solver (see \cite{Adamek:2015eda,Adamek:2016zes} for alternative approaches). In this Newtonian approach, the evolution of massless radiation species is usually ignored - although different methods to incorporate these effects have been proposed (see e.g. \cite{Fidler:2015npa,Fidler:2017ebh,Brandbyge:2016raj,Adamek:2017grt}). 

A simple fix to include these relativistic corrections in Newtonian simulations is the method of back-scaling. The initial conditions are calculated by rescaling the \emph{relativistic} but linear present-day power spectrum computed in an Einstein-Boltzmann solver back to the initial time using the scale-independent matter growth function. Then when evolving forwards from these initial conditions in Newtonian gravity, the relativistic power spectrum is recovered trivially at the present time, at least on the linear scales where the relativistic corrections are most important. It was shown in \cite{Fidler:2017ebh} that this method is in fact consistent with GR at all times and up to the very small scales if the results are interpreted using a remarkably simple GR-dictionary.

While the method to find initial conditions in cosmologies with massless neutrinos is well understood, incorporating massive neutrinos significantly complicates the problem. The reason is that massive neutrinos have a non-trivial impact on the evolution of the matter perturbations. On scales smaller than the neutrino free-streaming scale they suppress the formation of structures \cite{2013neco.bookL}, and depending on their mass they eventually start to accumulate around the most massive dark matter structures, leading to very rich dynamics on the smallest scales. 

One method to simulate the dynamics introduced by the massive neutrinos is adding an additional species to the non-linear N-body simulation, either represented by particles or as a fluid \cite{Adamek:2017uiq,Dakin:2017idt}. While these approaches are able to capture the complex neutrino dynamics, they are computationally significantly more expensive than ordinary CDM-only simulations.

Since back-scaling works extremely well for photons one may wonder if a similar method can be used also in the case of massive neutrinos and whether the rich dynamics can be introduced by starting the simulation from carefully designed initial conditions. One might for example try the usual back-scaling method in a massive neutrino cosmology or employ a back-scaling specifically designed for massive neutrinos such as presented in \cite{villaescusa_et_al}.

In this paper we aim to discuss the viability of such back-scaling approaches in cosmologies with massive neutrinos. We employ the Newtonian motion gauge framework, which allows the interpretation of Newtonian simulations in GR, ref. \cite{Fidler:2016tir}. This approach has been demonstrated to be consistent with GR in the weak-field limit in \cite{NLNM} and has recently been applied to multi-species systems such as a fluid of CDM plus baryons and massive neutrinos in \cite{Fidler:2018bkg}. This framework is particularly useful for comparing different methods for finding initial conditions as there exists one unique Nm gauge for any possible set of initial conditions and by studying this gauge we can understand if the method is self-consistent within the weak-field limit of GR and if a simple interpretation of the output exists.  

\subsection{Notation and Conventions}

We only consider scalar perturbations for simplicity. Assuming summation over repeated indices we have the metric line element

\be \dd s^2 = g_{\mu\nu} \,\dd x^\mu \dd x^\nu = g_{00}\, \dd \tau^2 + 2 g_{0i}\, \dd x^i \dd \tau + g_{ij} \,\dd x^i \dd x^j \,,
\ee
in an unspecified gauge with the following metric coefficients:

\begin{subequations}
\label{metric-potentials}
\begin{align}
  g_{00} &= -a^2 \left[ 1 + 2 A \right] \,, \\
  g_{0i} &=  -a^2  \hat\nab_i  B  \,,\\
  g_{ij} &= a^2 \left[ \delta_{ij} \left( 1 + 2 \HL \right) + 2 \left(  \hat\nab_i \hat\nab_j + \frac {\delta_{ij}}  {3} \right) \HT  \right] \,,
\end{align}
\end{subequations}
where $\hat\nab_i$ denotes the normalised gradient operator $\hat\nab_i \equiv -(-\nabla^2)^{-1/2} \nab_i$, known as the Riesz transform, that we use here to make the weak-field order of the perturbations explicit. We use the conformal time variable $\tau$, defined via $a\, \dd \tau = \dd t$, where $a=a(\tau)$ is the cosmological scale factor which evolves according to the Friedmann equations. 

The Einstein equations governing the evolution of the scalar perturbations $A$, $B$, $\HL$ and $\HT$ are sourced by the non-linear total stress-energy tensor, in space-time components 
\begin{subequations}
\label{energy-momentum}
\begin{align}
  T^0_{\phantom{0}0} &= -\sum_X \rho_X \equiv - \rho \,, \\
 T^0_{\phantom{0}i} &= \sum_X[\rho_X+ p_X] \hat\nab_i (  v_X -  B ) \equiv [\rho+ p] \hat\nab_i (  v -  B ) \,, \\ 
 T^i_{\phantom{i}j} &=  \sum_X \left[ p_X \delta^i_j  + \left(  \hat\nab^i \hat\nab_j + \frac {\delta^i_{j}} {3} \right) \Sigma_X  + [\rho_{X}+p_{X}] ( \hat\nab^i v_{X}) \hat\nab_j v_{X} \right] \nonumber \\ 
  & \equiv p \delta^i_j  + \left(  \hat\nab^i \hat\nab_j + \frac {\delta^i_{j}} {3} \right) \Sigma  + [\rho+p] (\hat\nab^i v) \hat\nab_j v \,,
\end{align}
\end{subequations}
where the sums over $X$ account for all of the relevant species in the universe, and $\rho_X$,  $p_X$, $\hat\nab_i v_X$ and $\Sigma_X$ are respectively the density, pressure, velocity and anisotropic stress of species $X$.

\subsection{Newtonian-Motion Gauges}

The Newtonian motion (Nm) gauge framework has been developed in \cite{Fidler:2016tir,Fidler:2017ebh,NLNM,Fidler:2018bkg}. It allows for the usage of unmodified Newtonian simulations to obtain a full GR solution including the effect of radiation on structure formation.
In \cite{Fidler:2016tir,Fidler:2017ebh}, the evolution of the relativistic species and their feedback on the non-linear matter evolution, as well as the relativistic space-time for standard Newtonian simulations were presented using linear CPT. This assumption has been relieved in \cite{NLNM}, where the Newtonian-motion gauge framework is discussed in the weak-field limit that remains valid until the very small scales of strong gravity. For in-depth presentations of the Newtonian-motion framework we refer to the aforementioned works. Here we briefly outline the essential ideas. 

GR introduces several terms that do not appear in the Newtonian equations of motion for matter. Specifically, the relativistic Euler equation which enforces momentum conservation includes GR corrections that can however be accurately computed in linear CPT since these corrections are relevant mostly on the large scales. The Nm gauges provide a framework in which these terms are directly absorbed in the definition of the coordinate system. Using the gauge freedom of GR a spatial gauge condition is found such that the relativistic Euler equation of matter is consistent with Newtonian theory. The resulting dynamically evolving coordinate system can be calculated using a linear Einstein-Boltzmann solver independent of the unmodified Newtonian simulation. By interpreting the particle positions obtained in the Newtonian simulation in these coordinates we obtain results that are consistent with GR in the weak-field limit.

It has been shown in \cite{Fidler:2016tir} that the temporal Poisson gauge condition $kB=\HT$ is a suitable choice for Nm gauges such that all metric potentials remain perturbatively small. With the comoving curvature perturbation $\zeta$\footnote{Within our metric conventions, the comoving curvature perturbation is $\zeta = \HL + \HT/3 - \Hc k^{-1}(v -B)$.}, the spatial gauge condition reads
\begin{align} \nonumber \label{eq:gauge-def}
\left(\partial_\tau +\Hc\right)&\dot{H}_{\rm T} \\ 
=& 4\pi G a^2 (\delta\rho_\gamma +\rho_m (\HT - 3\zeta + \delta M + \bar{M} \delta_m) + 4 \Hc  \rho_\gamma k^{-1}(v - k^{-1}\dot{H}_{\rm T}) + 2 \Sigma) - k T_\gamma \,,
\end{align}
where the subscript $\gamma$ denotes quantities related to the relativistic species evolved in the Einstein-Boltzmann solver and the subscript $m$ denotes perturbations of the massive species whose evolution is calculated non-linearly within the N-body simulation. Here, we have also introduced a mass modulation $M=\bar{M}+\delta M$ and relativistic corrections $T_\gamma$. The mass modulation is defined via the relation between the relativistic matter density and the counting density that is obtained from the Newtonian simulation via $\rho_m = (1 - 3\HL + M)\rho_{\rm counting}$. These terms account for the possibility that a species (such as massive neutrinos) transitions from the relativistic fluid to the massive one during the runtime of the simulation. In that case the density in the simulation still evolves according to an unmodified Vlasov-Poisson equation whereas the full matter density receives an additional contribution from the species that becomes non-relativistic. The mass modulation evolves according to
\be
\dot{\bar{M}} = (1 - \bar{M})(3\Hc + \frac{\dot{\bar{\rho}}_m}{\bar\rho_m}) \,,
\ee
and 
\begin{align} \nonumber
\delta \dot{M} + (3\Hc + \frac{\dot{\bar{\rho}}_m}{\bar\rho_m}) \delta M = \;
&3(3\Hc +\frac{\dot{\bar{\rho}}_m}{\bar\rho_m}) \HL 
- \dot{\bar{M}} \delta_m
+ 3 \bar{M} \dot{H}_{\rm L} \\
- &\frac{1 - \bar{M}}{\bar\rho_m}\left[\delta\dot{\rho}_\gamma + 4\Hc\delta\rho_\gamma + 4\Hc\bar{\rho}_\gamma\dot{H}_{\rm L} - \frac 4 3 \bar{\rho}_\gamma k \hat\nab^i\hat\nab_i v_\gamma\right] \,.
\end{align}
In the simplest possible case of a pure CDM Universe free from radiation, the mass modulation and all terms related to the relativistic fluid cancel. We then identify $\HT=3\zeta$ as a trivial equilibrium solution of the above gauge defenition \ref{eq:gauge-def}. We define the N-boisson gauge with $\HT = 3\zeta$ and $k B =  3\dot{\zeta}$ as a the Nm gauge suitable for such CDM-only simulations. This gauge is closely related to the GR dictionary described in \cite{Chisari:2011iq}.

\section{Initial Conditions}

The gauge condition for Newtonian-motion gauges can be conveniently formulated as a second order differential equation for $\HT$ (see equation \ref{eq:gauge-def}). This leaves two residual degrees of freedom corresponding to the initial conditions for $\HT$ and its derivative that may be chosen freely. These correspond directly to the initial Newtonian density and velocity such that a unique Newtonian-motion gauge can be found for any set of initial conditions used in the Newtonian simulation. This allows us to test wether certain proposed methods to initialise simulations are self-consistent from a full relativistic point of view and in which coordinates these should be interpreted. 

\subsection{Vanilla Back-Scaling}\label{sec:vanilla}

\begin{figure}[htbp]
	\label{backscaling-sketch}
	\centering
	\includegraphics[width=0.95\textwidth]{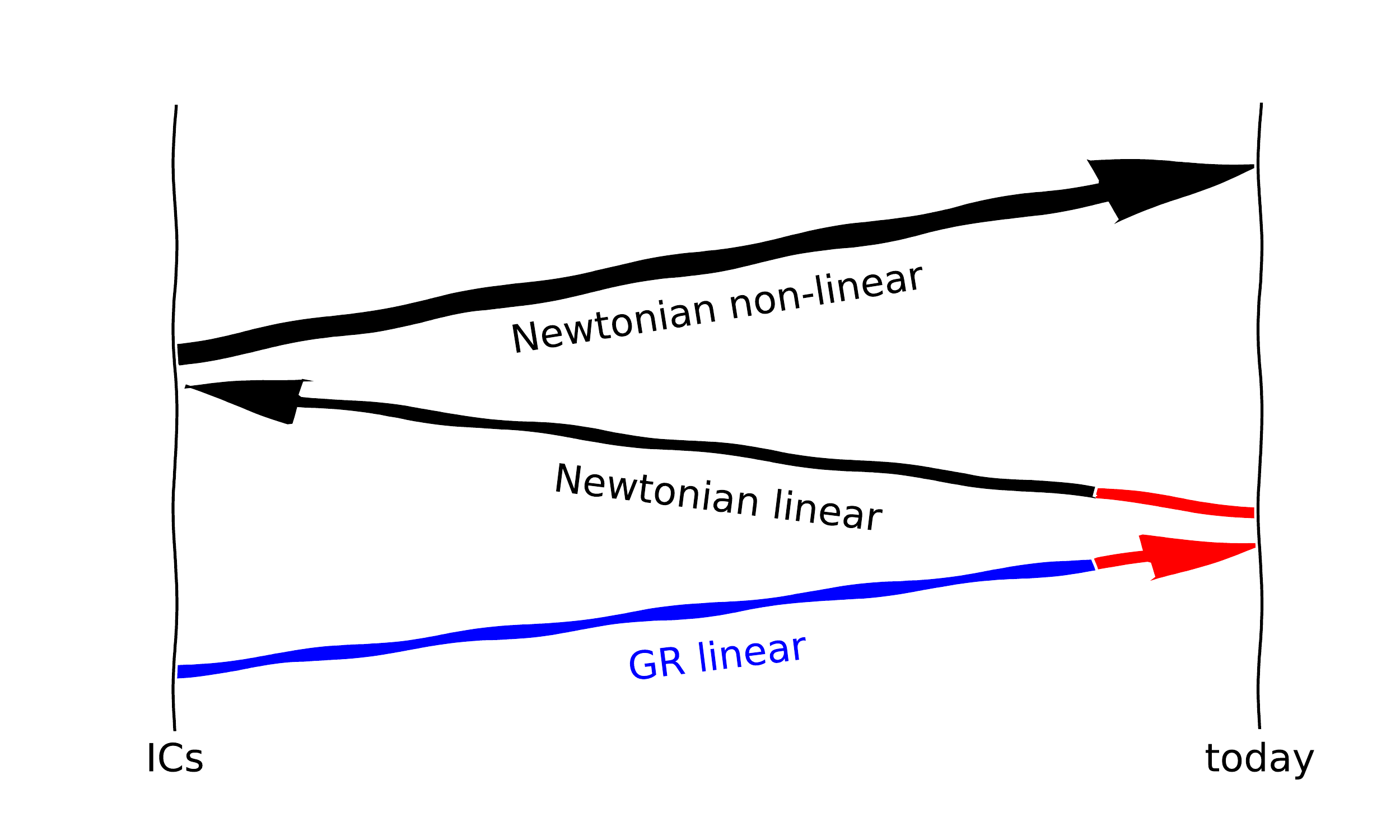}
	\caption{A sketch visualising the basic idea of back-scaling. The blue arrow describes a relativistic evolution while black arrows are Newtonian. Thin arrows represent a linear approximation that becomes inaccurate on the small scales towards the end of the evolution. We indicate this breakdown of perturbation theory by colouring the arrows red.}
\end{figure}

The commonly applied method of back-scaling employs the relativistic but linear present-day power spectrum and then solves the linear Newtonian dynamics backwards until the initial time. The idea is illustrated in Figure \ref{backscaling-sketch} where the blue line represents the relativistic but linear evolution in a Boltzmann code. The result is then rescaled back to the initial time using the linear Newtonian theory following the thin black arrow. Finally from these initial conditions the N-body simulation solves a non-linear but Newtonian evolution illustrated by the thick black arrow.
This method is routinely applied to incorporate relativistic and radiation corrections in Newtonian N-body simulations and has been shown to have many favourable properties in \cite{Fidler:2017ebh}, especially the existence of a simple GR dictionary that links the relativistic perturbations to the Newtonian simulation. 

The method works because of two simple cancellations between the forwards and backwards evolution. On the large scales the two black arrows are identical since non-linearities remain small. The entire back-scaling method thus reproduces the blue arrow describing the relativistic evolution. On the other hand, for the small scales the relativistic evolution is almost identical to the Newtonian one so that both linear (thin) arrows cancel and we obtain the non-linear result, illustrated by the thick black arrow, starting from unmodified initial conditions.
This cancellation is crucial since on the small scales the present day linear solution cannot be trusted since it is well beyond its range of validity, illustrated by the red parts of the arrows. The initial conditions should not depend on such an unphysical quantity and this is guaranteed by the cancellation among the two linear evolutions. In that way back-scaling modifies the initial perturbations only on the large scales, including the impact of GR and radiation, while the small scales where non-linear corrections are important are not affected.

However this does no longer hold when considering massive neutrinos. Neutrinos do modify the evolution of matter compared to a massive Newtonian analysis. The cancellation between the linear evolutions (thin arrows) is lost and the back-scaled initial conditions are modified on all scales and now depend explicitly on the value of the unphysical linear present day matter density. While we still trivially recover the linear matter power spectrum on the larger scales (the cancellation between the two Newtonian arrows still holds), we have potentially introduced a significant bias in our non-linear evolution. x
  
Furthermore the interpretation of the simulation may become more complicated. In the massless case we know that we can interpret the Newtonian simulation in the N-boisson gauge, at least at sufficiently late times. This is because our condition imposed on the densities implies a simple condition also for the metric potentials. However, for massive neutrinos we have left the range of validity for linear perturbation theory and our condition on the density at the present time no longer sets a consistent condition for the metric. 

\begin{figure}[htbp]
	\label{back-scaling-plot}
	\centering
	\includegraphics[width=0.95\textwidth]{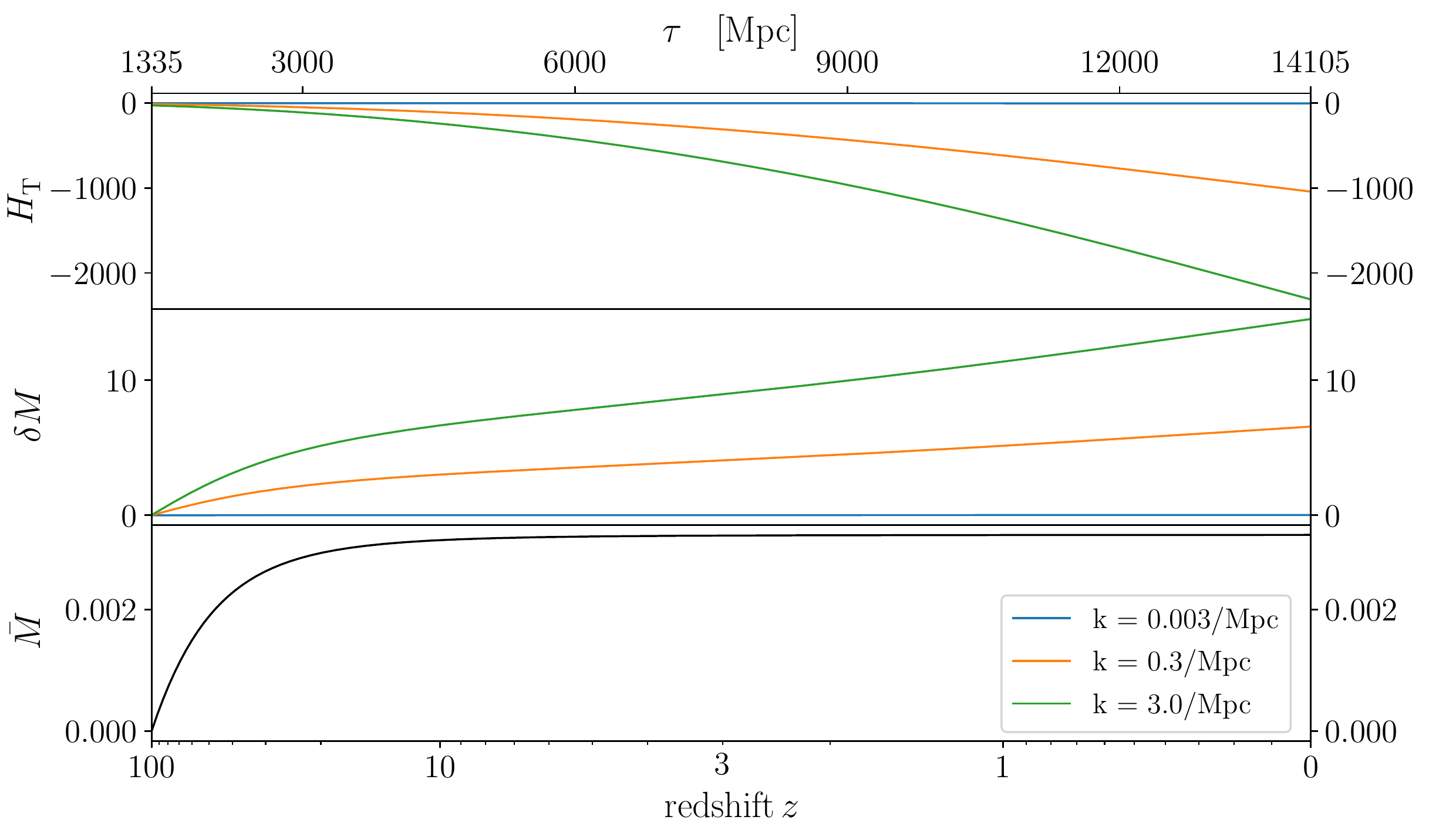}
	\caption{The metric potential $\HT$ and the mass modulation $M$ starting from conventional back-scaling initial conditions for a combined neutrino mass of $170$ meV. While the mass modulation remains small, the metric potential $\HT$ grows quickly on the small scales and reaches values that are in conflict with the weak field assumptions.}
\end{figure}

The Nm gauge framework allows us to compute the metric corresponding to back-scaling initial conditions, shown in figure \ref{back-scaling-plot} for a combined neutrino mass of $170$ meV.
We find that the metric perturbations become dangerously large on the small scales. Our weak-field assumptions may be broken implying that additional non-linear corrections introduced by our choice of initial conditions need to be accounted for at the level of the metric. The metric also does evolve at the late times, requiring a complex GR dictionary to interpret such a simulation. 

\subsection{Tailored Back-Scaling for Neutrino Cosmologies}

So far we have considered only the simplest model of back-scaling, where the present day power spectrum is rescaled with the linear matter growth function as in the massless neutrino case. In \cite{villaescusa_et_al} a more complex method of back-scaling was developed specifically for massive neutrinos. The authors account for the different evolution of the neutrinos compared to the other matter components by employing a two-fluid approach for the growth function.

\begin{figure}[htbp]
	\label{villaescusa-plot}
	\centering
	\includegraphics[width=0.95\textwidth]{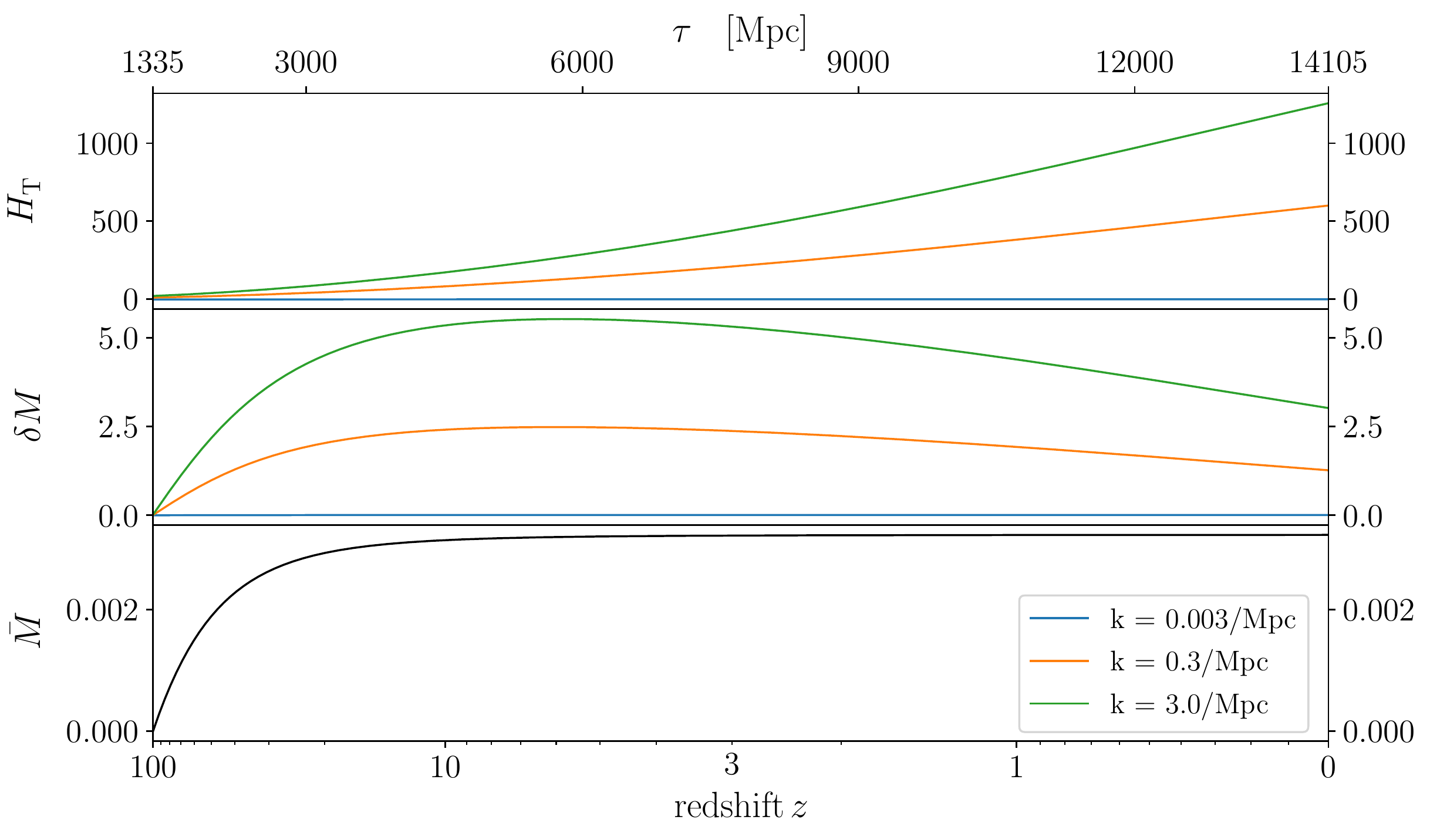}
	\caption{The metric potential $\HT$ and the mass modulation $M$ in the approach of \cite{villaescusa_et_al} for a combined neutrino mass of $170$ meV. Again we find that the metric potentials on the small scales become very large and are time-dependent.}
\end{figure}

We generate initial conditions according to this method using the code {\sc reps} (see \cite{initial_conditions_generator_zennaro}) and construct the corresponding metric in weak-field relativity. Again we find that the metric becomes large and evolves non-trivially on the smaller scales as shown in Figure \ref{villaescusa-plot}. 
It turns out that $\HT$ now evolves in the opposite direction and remains a bit smaller showing that these more advanced initial conditions do improve the back-scaling method, but they seem to overcompensate for the problems of conventional back-scaling. However, the initial conditions from \cite{villaescusa_et_al} also introduce a small correction on the large scales where the linear result is no longer exactly reproduced in the Newtonian simulation. In the logic of back-scaling this is a necessary price to pay. The method includes the Neutrino physics more accurately in the growth-function (thin black arrow), improving the cancelation between the two thin lines in figure \ref{backscaling-sketch}. But since the thick black line is still realised by an unmodified Newtonian simulation, it now does no longer cancel with the thin black arrow in the linear regime. As a consequence the relativistic solution can no longer be reproduced exactly. This however is only a small inconvenience as this mismatch can easily be absorbed into the metric when using the Newtonian motion gauge approach. 

\subsection{Backwards, not Back-Scaled}
In our last paper \cite{Fidler:2018bkg} we have introduced so called \emph{backwards} initial conditions. 
In this method we do not employ the unphysical linear densities at the present time to define the initial conditions but instead use the metric potential $\HT$ which remains accurate in linear theory up to the very small scales and late times. Instead of fixing $\delta_m$, we demand that $\HT$ takes on a simple value at the final time, i.e. $\HT = 3\zeta$, corresponding to the N-boisson gauge. Then the gauge condition is solved backwards to find the weak-field consistent initial metric and from there the initial values for $\delta_m$ in linear theory. In this way we completely avoid the problems introduced by using a linear approximation for a non-linear small-scale perturbation and all of our calculations remain consistent to weak-field precision. 

In the case of massless neutrinos this turns out to be identical to back-scaling since the initial conditions in this case do not depend on the small-scale present day densities anyway due to the cancellation between the linear Newtonian and relativistic theories. However if we add massive neutrinos the corresponding initial conditions start to differ from conventional back-scaling.

Solving the evolution backwards is a complex problem and instead we employ a shooting method which iteratively finds initial conditions which approximate the desired final result.
\begin{figure}[htbp]
	\label{backwards-plot}
	\centering
	\includegraphics[width=0.95\textwidth]{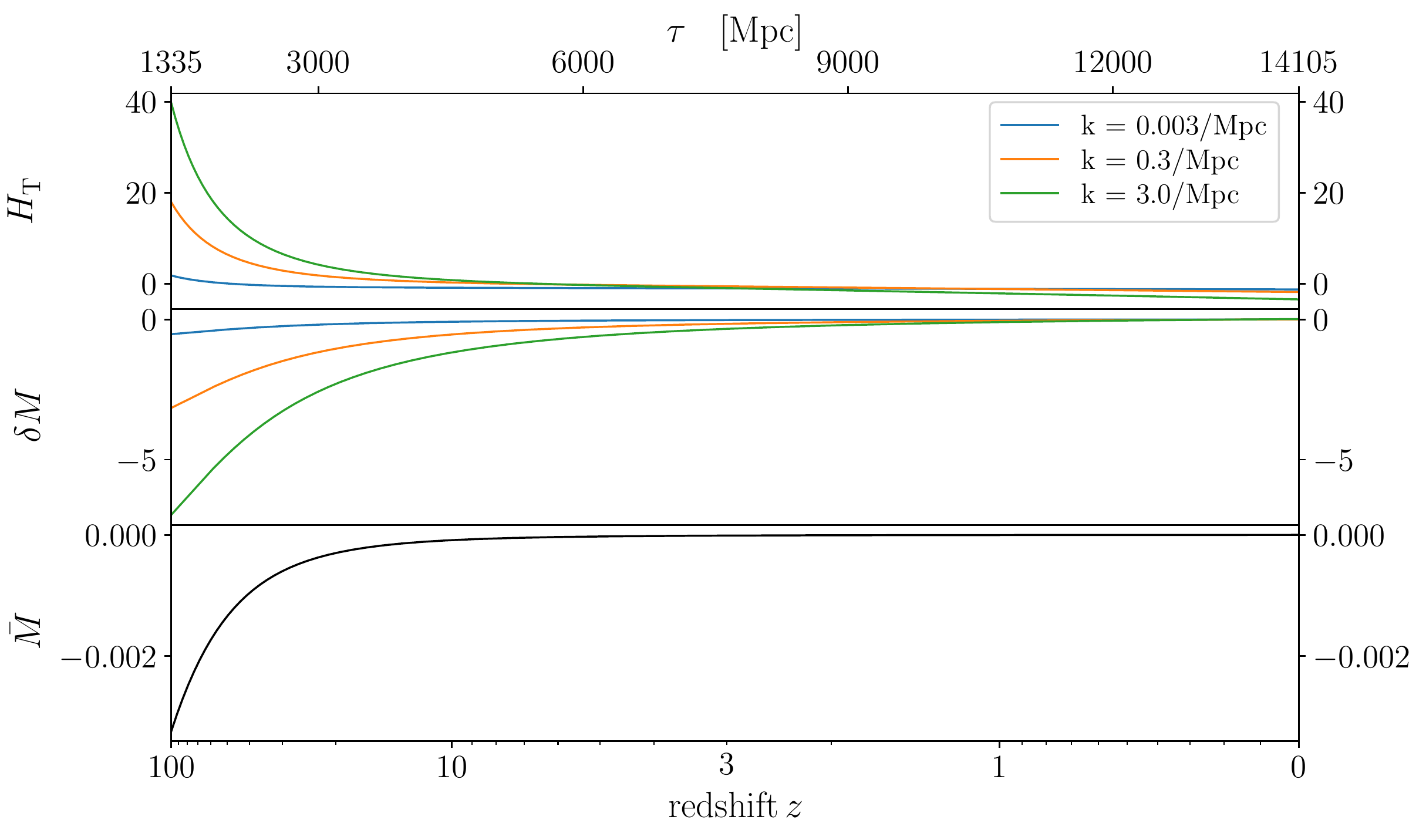}
	\caption{The metric potential $\HT$ and the mass modulation $M$ in the backwards approach for a combined neutrino mass of $170$ meV. The metric remains perturbatively small and assumes an almost constant late time limit. The residual evolution results from our numerical value not exactly finding the optimum and from the small residual impacts of neutrinos at the very late times.  }
\end{figure}
The metric we derive for this type of initial conditions is shown in Figure \ref{backwards-plot}. The metric perturbations remain almost two orders of magnitude smaller than in the case of back-scaling, showing that we successfully absorbed the non-linear corrections into the initial conditions. Furthermore the late-time metric becomes well behaved and approaches a simple limit which allows for an easy analysis of the data. Note that the metric does not become perfectly flat at the late time since we use a numerical method.

\section{Conclusions}
We have investigated various methods for setting initial conditions for Newtonian N-body simulations in cosmologies with massive neutrinos. 
Many simulations employ back-scaling initial conditions to include the impact of relativistic species and GR due to the simplicity and robustness of this method. Using the Newtonian motion gauge framework we study the suitability of this approach when including massive neutrinos and construct the coordinate system in which such simulations can be interpreted in agreement with weak-field relativity. 

While back-scaling methods are very successful for massless neutrinos, we show that they fail in the case of massive neutrinos. We identify the fundamental problem of such initial conditions as their explicit dependence on the present day matter power spectrum from linear theory, which is completely outside of the regime of validity. In the massless case this problem was avoided since the initial conditions decouple from the late-time density on the small scales due to a cancellation between the relativistic forwards evolution in the Boltzmann code and the Newtonian back-scaling with the linear growth function. Neutrinos however modify the rate of structure formation in the Universe and the back-scaling initial conditions now depend explicitly on the unphysical linear matter overdensity. 

When constructing the weak-field metric corresponding to such initial conditions we find that the metric potentials need to be very large in order to compensate for the mismatch in the initial conditions and accurately include the neutrino dynamics. Consequently the interpretation of such a simulation becomes non-trivial requiring a time-dependent GR dictionary. Even more problematically, the weak-field assumption is broken hinting at leading order corrections being missed. Since the non-linear evolution mixes different scales, this may contaminate the output on all scales.

We found these problems not only for ordinary back-scaling, but also for more complex methods that are specifically designed to include the non-trivial evolution of neutrinos \cite{villaescusa_et_al}. The reason is that these methods still depend on the present day linear matter power spectrum. Furthermore back-scaling relies on two cancelations as argued in section \ref{sec:vanilla}. By trying to restore the small-scale cancellation between the relativistic and back-scaled evolution, the second large-scale cancellation between the Newtonian simulation and the back-scaling is lost.  

We show that this problem can be avoided by specifying more carefully which conditions can consistently be enforced at the final time. Instead of using a condition based on the present day density, we demand that the present day metric remains simple. This is a suitable condition since $\HT$ is shielded from non-linear corrections and remains well described in linear theory until the present time. The resulting initial conditions are then labeled as backwards and not back-scaled. 

We find that these initial conditions have properties that are very similar to ordinary back-scaling in a massless neutrino case. The metric remains small at all times and a simple late-time limit for the metric exists. The main difference is the computation of these initial conditions, which is significantly more complex than applying a rescaling to the present day power spectrum. Nevertheless the task of finding suitable initial conditions is computationally fast compared to a typical N-body simulation.

The backwards initial conditions can be used in combination with any ordinary Newtonian N-body simulation and absorb the complex neutrino dynamics into a non-linear part encoded in the initial conditions and a simple linear dictionary to interpret the output. The implementation and numerical analysis of this method for Newtonian N-body simulations will be studied in future work.

\bibliographystyle{JHEP}
\bibliography{references}

\providecommand{\href}[2]{#2}\begingroup\raggedright\begin{thebibliography}{10}

\bibitem{Kodama:1985bj}
H.~Kodama and M.~Sasaki, \emph{{Cosmological Perturbation Theory}},
  \href{http://dx.doi.org/10.1143/PTPS.78.1}{\emph{Prog. Theor. Phys. Suppl.}
  {\bf 78} (1984) 1--166}.

\bibitem{Malik:2008im}
K.~A. Malik and D.~Wands, \emph{{Cosmological perturbations}},
  \href{http://dx.doi.org/10.1016/j.physrep.2009.03.001}{\emph{Phys. Rept.}
  {\bf 475} (2009) 1--51}, [\href{http://arxiv.org/abs/0809.4944}{{\tt
  0809.4944}}].

\bibitem{Villa:2015ppa}
E.~Villa and C.~Rampf, \emph{{Relativistic perturbations in $\Lambda$CDM:
  Eulerian \& Lagrangian approaches}},
  \href{http://dx.doi.org/10.1088/1475-7516/2016/01/030}{\emph{JCAP} {\bf 1601}
  (2016) 030}, [\href{http://arxiv.org/abs/1505.04782}{{\tt 1505.04782}}].

\bibitem{class2}
D.~Blas, J.~Lesgourgues and T.~Tram, \emph{{The Cosmic Linear Anisotropy
  Solving System (CLASS) II: Approximation schemes}},
  \href{http://dx.doi.org/10.1088/1475-7516/2011/07/034}{\emph{JCAP} {\bf 1107}
  (2011) 034}, [\href{http://arxiv.org/abs/1104.2933}{{\tt 1104.2933}}].

\bibitem{Lewis:1999bs}
A.~Lewis, A.~Challinor and A.~Lasenby, \emph{{Efficient computation of CMB
  anisotropies in closed FRW models}},
  \href{http://dx.doi.org/10.1086/309179}{\emph{Astrophys. J.} {\bf 538} (2000)
  473--476}, [\href{http://arxiv.org/abs/astro-ph/9911177}{{\tt
  astro-ph/9911177}}].

\bibitem{Teyssier:2001cp}
R.~Teyssier, \emph{{Cosmological hydrodynamics with adaptive mesh refinement: a
  new high resolution code called ramses}},
  \href{http://dx.doi.org/10.1051/0004-6361:20011817}{\emph{Astron. Astrophys.}
  {\bf 385} (2002) 337--364},
  [\href{http://arxiv.org/abs/astro-ph/0111367}{{\tt astro-ph/0111367}}].

\bibitem{Springel:2005mi}
V.~Springel, \emph{{The Cosmological simulation code GADGET-2}},
  \href{http://dx.doi.org/10.1111/j.1365-2966.2005.09655.x}{\emph{Mon. Not.
  Roy. Astron. Soc.} {\bf 364} (2005) 1105--1134},
  [\href{http://arxiv.org/abs/astro-ph/0505010}{{\tt astro-ph/0505010}}].

\bibitem{Hahn:2015sia}
O.~Hahn and R.~E. Angulo, \emph{{An adaptively refined phase-space element
  method for cosmological simulations and collisionless dynamics}},
  \href{http://dx.doi.org/10.1093/mnras/stv2304}{\emph{Mon. Not. Roy. Astron.
  Soc.} {\bf 455} (2016) 1115--1133},
  [\href{http://arxiv.org/abs/1501.01959}{{\tt 1501.01959}}].

\bibitem{Adamek:2015eda}
J.~Adamek, D.~Daverio, R.~Durrer and M.~Kunz, \emph{{General relativity and
  cosmic structure formation}},
  \href{http://dx.doi.org/10.1038/nphys3673}{\emph{Nature Phys.} {\bf 12}
  (2016) 346--349}, [\href{http://arxiv.org/abs/1509.01699}{{\tt 1509.01699}}].

\bibitem{Adamek:2016zes}
J.~Adamek, D.~Daverio, R.~Durrer and M.~Kunz, \emph{{gevolution: a cosmological
  N-body code based on General Relativity}},
  \href{http://dx.doi.org/10.1088/1475-7516/2016/07/053}{\emph{JCAP} {\bf 1607}
  (2016) 053}, [\href{http://arxiv.org/abs/1604.06065}{{\tt 1604.06065}}].

\bibitem{Fidler:2015npa}
C.~Fidler, C.~Rampf, T.~Tram, R.~Crittenden, K.~Koyama and D.~Wands,
  \emph{{General relativistic corrections to $N$-body simulations and the
  Zel'dovich approximation}},
  \href{http://dx.doi.org/10.1103/PhysRevD.92.123517}{\emph{Phys. Rev.} {\bf
  D92} (2015) 123517}, [\href{http://arxiv.org/abs/1505.04756}{{\tt
  1505.04756}}].

\bibitem{Fidler:2017ebh}
C.~Fidler, T.~Tram, C.~Rampf, R.~Crittenden, K.~Koyama and D.~Wands,
  \emph{{Relativistic initial conditions for N-body simulations}},
  \href{http://dx.doi.org/10.1088/1475-7516/2017/06/043}{\emph{JCAP} {\bf 1706}
  (2017) 043}, [\href{http://arxiv.org/abs/1702.03221}{{\tt 1702.03221}}].

\bibitem{Brandbyge:2016raj}
J.~Brandbyge, C.~Rampf, T.~Tram, F.~Leclercq, C.~Fidler and S.~Hannestad,
  \emph{{Cosmological $N$-body simulations including radiation perturbations}},
  \href{http://dx.doi.org/10.1093/mnrasl/slw235}{\emph{Mon. Not. Roy. Astron.
  Soc.} {\bf 466} (2017) L68--L72},
  [\href{http://arxiv.org/abs/1610.04236}{{\tt 1610.04236}}].

\bibitem{Adamek:2017grt}
J.~Adamek, J.~Brandbyge, C.~Fidler, S.~Hannestad, C.~Rampf and T.~Tram,
  \emph{{The effect of early radiation in N-body simulations of cosmic
  structure formation}},
  \href{http://dx.doi.org/10.1093/mnras/stx1157}{\emph{Mon. Not. Roy. Astron.
  Soc.} (2017) }, [\href{http://arxiv.org/abs/1703.08585}{{\tt 1703.08585}}].

\bibitem{2013neco.bookL}
J.~{Lesgourgues}, G.~{Mangano}, G.~{Miele} and S.~{Pastor}, \emph{{Neutrino
  Cosmology}}.
\newblock Feb., 2013.

\bibitem{Adamek:2017uiq}
J.~Adamek, R.~Durrer and M.~Kunz, \emph{{Relativistic N-body simulations with
  massive neutrinos}},
  \href{http://dx.doi.org/10.1088/1475-7516/2017/11/004}{\emph{JCAP} {\bf 1711}
  (2017) 004}, [\href{http://arxiv.org/abs/1707.06938}{{\tt 1707.06938}}].

\bibitem{Dakin:2017idt}
J.~Dakin, J.~Brandbyge, S.~Hannestad, T.~Haugbølle and T.~Tram,
  \emph{{$\nu$CO$N$CEPT: Cosmological neutrino simulations from the non-linear
  Boltzmann hierarchy}},  \href{http://arxiv.org/abs/1712.03944}{{\tt
  1712.03944}}.

\bibitem{villaescusa_et_al}
M.~Zennaro, J.~Bel, F.~Villaescusa-Navarro, C.~Carbone, E.~Sefusatti and
  L.~Guzzo, \emph{Initial conditions for accurate n-body simulations of massive
  neutrino cosmologies},
  \href{http://dx.doi.org/10.1093/mnras/stw3340}{\emph{Monthly Notices of the
  Royal Astronomical Society} {\bf 466} (2017) 3244--3258}.

\bibitem{Fidler:2016tir}
C.~Fidler, T.~Tram, C.~Rampf, R.~Crittenden, K.~Koyama and D.~Wands,
  \emph{{Relativistic Interpretation of Newtonian Simulations for Cosmic
  Structure Formation}},
  \href{http://dx.doi.org/10.1088/1475-7516/2016/09/031}{\emph{JCAP} {\bf 1609}
  (2016) 031}, [\href{http://arxiv.org/abs/1606.05588}{{\tt 1606.05588}}].

\bibitem{NLNM}
C.~Fidler, T.~Tram, C.~Rampf, R.~Crittenden, K.~Koyama and D.~Wands,
  \emph{{General relativistic weak-field limit and Newtonian N-body
  simulations}},
  \href{http://dx.doi.org/10.1088/1475-7516/2017/12/022}{\emph{JCAP} {\bf 1712}
  (2017) 022}, [\href{http://arxiv.org/abs/1708.07769}{{\tt 1708.07769}}].

\bibitem{Fidler:2018bkg}
C.~Fidler, A.~Kleinjohann, T.~Tram, C.~Rampf and K.~Koyama, \emph{{A new
  approach to cosmological structure formation with massive neutrinos}},
  \href{http://arxiv.org/abs/1807.03701}{{\tt 1807.03701}}.

\bibitem{Chisari:2011iq}
N.~E. Chisari and M.~Zaldarriaga, \emph{{Connection between Newtonian
  simulations and general relativity}},
  \href{http://dx.doi.org/10.1103/PhysRevD.84.089901,
  10.1103/PhysRevD.83.123505}{\emph{Phys. Rev.} {\bf D83} (2011) 123505},
  [\href{http://arxiv.org/abs/1101.3555}{{\tt 1101.3555}}].

\bibitem{initial_conditions_generator_zennaro}
matteozennaro, \emph{matteozennaro/reps: Reps - version jan 2016},  Jan., 2017.
\newblock 10.5281/zenodo.259660.

\end{thebibliography}\endgroup

\end{document}